\documentclass[s]{iucr}              % DO NOT DELETE THIS LINE

\usepackage{epsfig}
\usepackage{amssymb}
\usepackage{amsmath}
\usepackage{textcomp}
\usepackage{ifthen}

     \paperprodcode{a000000}      % Replace with production code if known
     \paperref{xx9999}            % Replace xx9999 with reference code if known
     \papertype{FA}               % Indicate type of article
     \paperlang{english}          % Can be english, french, german or russian
     \journalcode{S}              % Indicate the journal to which submitted
     \journalyr{2003}
     \journaliss{1}
     \journalvol{59}
     \journalfirstpage{000}
     \journallastpage{000}
     \journalreceived{0 XXXXXXX 0000}
     \journalaccepted{0 XXXXXXX 0000}
     \journalonline{0 XXXXXXX 0000}

\begin{document}                  % DO NOT DELETE THIS LINE

     %-------------------------------------------------------------------------
     % The introductory (header) part of the paper
     %-------------------------------------------------------------------------

     % The title of the paper. Use \shorttitle to indicate an abbreviated title
     % for use in running heads (you will need to uncomment it).

\title{Development of a two-dimensional virtual pixel
X-ray imaging detector for time-resolved structure research}
\shorttitle{2D ViP detector development for time-resolved
experiments}

     % Authors' names and addresses. Use \cauthor for the main (contact) author.
     % Use \author for all other authors. Use \aff for authors' affiliations.
     % Use lower-case letters in square brackets to link authors to their
     % affiliations; if there is only one affiliation address, remove the [a].

\cauthor[a]{Andre}{Orthen}{orthen@alwa02.physik.uni-siegen.de}{}
\author[a]{Hendrik}{Wagner}
\author[a]{Sorin}{Martoiu}
\author[b]{Heinz}{Amenitsch}
\author[c]{Sigrid}{Bernstorff}
\author[a]{Hans-J\"urgen}{Besch}
\author[c]{Ralf-Hendrik}{Menk}
\author[a]{Kivan\c{c}}{Nurdan}
\author[b]{Michael}{Rappolt}
\author[a]{Albert Heinrich}{Walenta}
\author[a]{Ulrich}{Werthenbach}

\aff[a]{Universit\"at Siegen, Fachbereich Physik,
Emmy-Noether-Campus, Walter-Flex-Str. 3, D-57072 Siegen,
\country{Germany}} \aff[b]{Institute of Biophysics and X-Ray
Structure Research, Austrian Academy of Sciences, Schmiedlstr. 6,
A-8042 Graz, \country{Austria}} \aff[c]{Sincrotrone Trieste, S.S.
14, km 163.5, Basovizza, I-34012 Trieste, \country{Italy}}

     % Use \shortauthor to indicate an abbreviated author list for use in
     % running heads (you will need to uncomment it).

\shortauthor{Orthen, Wagner, Martoiu \emph{et. al}}

     % Use \vita if required to give biographical details (for authors of
     % invited review papers only). Uncomment it.

%\vita{Author's biography}

     % Keywords (required for Journal of Synchrotron Radiation only)
     % Use the \keyword macro for each word or phrase, e.g.
     % \keyword{X-ray diffraction}\keyword{muscle}


     % PDB and NDB reference codes for structures referenced in the article and
     % deposited with the Protein Data Bank and Nucleic Acids Database (Acta
     % Crystallographica Section D). Repeat for each separate structure e.g
     % \PDBref[dethiobiotin synthetase]{1byi} \NDBref[d(G$_4$CGC$_4$)]{ad0002}

%\PDBref[optional name]{refcode}
%\NDBref[optional name]{refcode}

\maketitle                        % DO NOT DELETE THIS LINE

\begin{synopsis}
%Development of a two-dimensional gaseous X-ray imaging detector
%for time-resolved structure research.
\end{synopsis}

\begin{abstract}
An interpolating two-dimensional X-ray imaging detector based on a
single photon counter with gas amplification by GEM (gas electron
multiplier) structures is presented. The detector system can be
used for time-resolved structure research down to the $\mathrm{\mu
s}$-time domain. The prototype detector has been tested at the
SAXS (small angle X-ray scattering) beamline at ELETTRA
synchrotron light source with a beam energy of $8\,\mathrm{keV}$.
The imaging performance is examined with apertures and standard
diffraction targets. Finally, the application in a time-resolved
lipid temperature jump experiment is presented.
\end{abstract}

     %-------------------------------------------------------------------------
     % The main body of the paper
     %-------------------------------------------------------------------------
     % Now enter the text of the document in multiple \section's, \subsection's
     % and \subsubsection's as required.

\section{Introduction}

Modern synchrotron radiation facilities are able to provide
extremely high photon fluxes with an enormous brilliance, which
opens up new research fields in many different areas. One of these
branches are fast time-resolved studies. Of special interest are
for example dynamical processes in material science \cite{Clery}
like phase transitions, polymerisation or deformations under
stress. In the chemical or biological domain for example muscle
contraction \cite{Squire,Wakabayashi,Holmes,Piazzesi} or lipid
phase transitions \cite{Rapp1992,Seddon} come into the focus of
interest.

However, the existing X-ray detectors do not offer the
time-resolutions needed or are often not capable of dealing with
the enormous photon rates produced by insertion devices like
wigglers, undulators or in future by free electron lasers, and are
the bottleneck of state of the art beamlines. To find a solution
for this mismatch one has to improve old detector concepts or to
introduce new approaches.

Detectors in general can be classified into two categories:
integrating detectors and single photon counters. The performance
of integrating devices like CCDs (charge coupled devices), image
plates or X-ray films, is getting more favourable at high photon
flux, since the relative noise contribution is decreasing. The
noise contributions can be attributed to noise caused by
fluctuations of the measured quantity related to the input
(\emph{intrinsic noise}), to \emph{dark noise} and to
\emph{readout noise}. The readout noise increases in proportion to
the readout frame rate, whereby in most CCDs this noise
contribution is much larger than the dark noise. Time-resolution
is thus competing with intensity precision; generally, the
time-resolution is limited to keep the readout noise contribution
relatively small. At some value of input flux the integrator
saturates and can not add up more until the content is reset. The
dynamic range of integrating devices, which is typically
$10^4$--$10^5$, is thus limited by noise and saturation level.

In contrast to integrating devices counting detectors like MWPCs
(multi wire proportional chambers) or MSGCs (micro strip gas
chambers) are not sensitive to noise contributions, described
above, as long as this noise level is below the threshold,
triggering the counter. However, one single event implies a dead
time, for instance caused by the length of the produced detector
signal and the electronic processing, e.g. the readout. Due to
this dead time the performance of counting detectors drops at high
rates and dead time corrections have to be applied to correct the
measured intensities. Especially when counting detectors are used
in synchrotron radiation facilities, these corrections can become
complicated due to the intrinsic time structure of the photon beam
\cite{Bateman}. The dynamic range of photon counting devices,
typically reaching values of up to $10^6$, is limited downwards by
the number of counts due to false triggers (noise triggers) and
upwards by the (inverse) dead time. To prevent the detector from
too high photon rates one has to attenuate the primary beam, and
thus the full power of modern synchrotrons is wasted. Due to the
low noise level, the high dynamic range and the good
time-resolution, which can be approximately in the order of
magnitude of the dead time or even better, single photon counting
detectors, e.g. one-dimensional delay-line detectors
\cite{Gabriel}, are often used for example in SAXS imaging, where
only small parallax is obtained due to the small diffraction
angles.

However, \emph{the} unique detector for every application is not
existent and compromises have to be found, optimising the detector
for the particular application. Although many efforts have been
started to increase the readout speed of integrating CCD cameras,
e.g. \cite{Tipnis}, the time-resolution is still limited to about
$1\,\mathrm{ms}$. Despite their good spatial resolution and their
large number of pixels CCDs also suffer from the limited intensity
precision of roughly $1\,\%$ \cite{Kocsis} and the limited dynamic
range. On the other hand, single photon counters can still not
deal with the enormous rates offered by synchrotron facilities.

There are new ideas to find solutions to this drawback e.g. with
semiconducting pixel array detectors \cite{Datte,Rossi,Renzi}.
With the introduction of micro pattern gaseous devices like MSGCs
\cite{Oed} or GEMs \cite{Sauli} several new approaches have been
carried out building faster gaseous detectors, e.g. the RAPID
(refined ADC per input detector) system
\cite{Lewis1994,Lewis1997,Lewis2000} or a GEM pixel detector
\cite{Bellazzini}, which replace the old detector era of MWPCs. An
overview of micro pattern gaseous detectors can be found for
example in \cite{Sauli1999}.

The two-dimensional single photon counting gaseous detector for
X-rays in the energy range of $5$--$25\,\mathrm{keV}$, described
in this work, represents a concept which is very flexible and can
therefore easily be adapted to special requirements. It offers
time-resolutions in the sub-microsecond range and is thus
qualified for time-resolved SAXS-imaging.

\section{Detector setup}

In this section we briefly describe the working principle and the
setup of the detector. Detailed information about the detector can
be found elsewhere
\cite{Besch,Sarvestani1998b,Wagner2003,Orthen2003b}. A review of a
previous, similar detector with smaller sensitive detection area
and a less sophisticated position reconstruction, with a different
gas gain structure and with slower electronics and readout has
also been published earlier \cite{Sarvestani1999b}.

\begin{figure}
  \includegraphics[width=8cm]{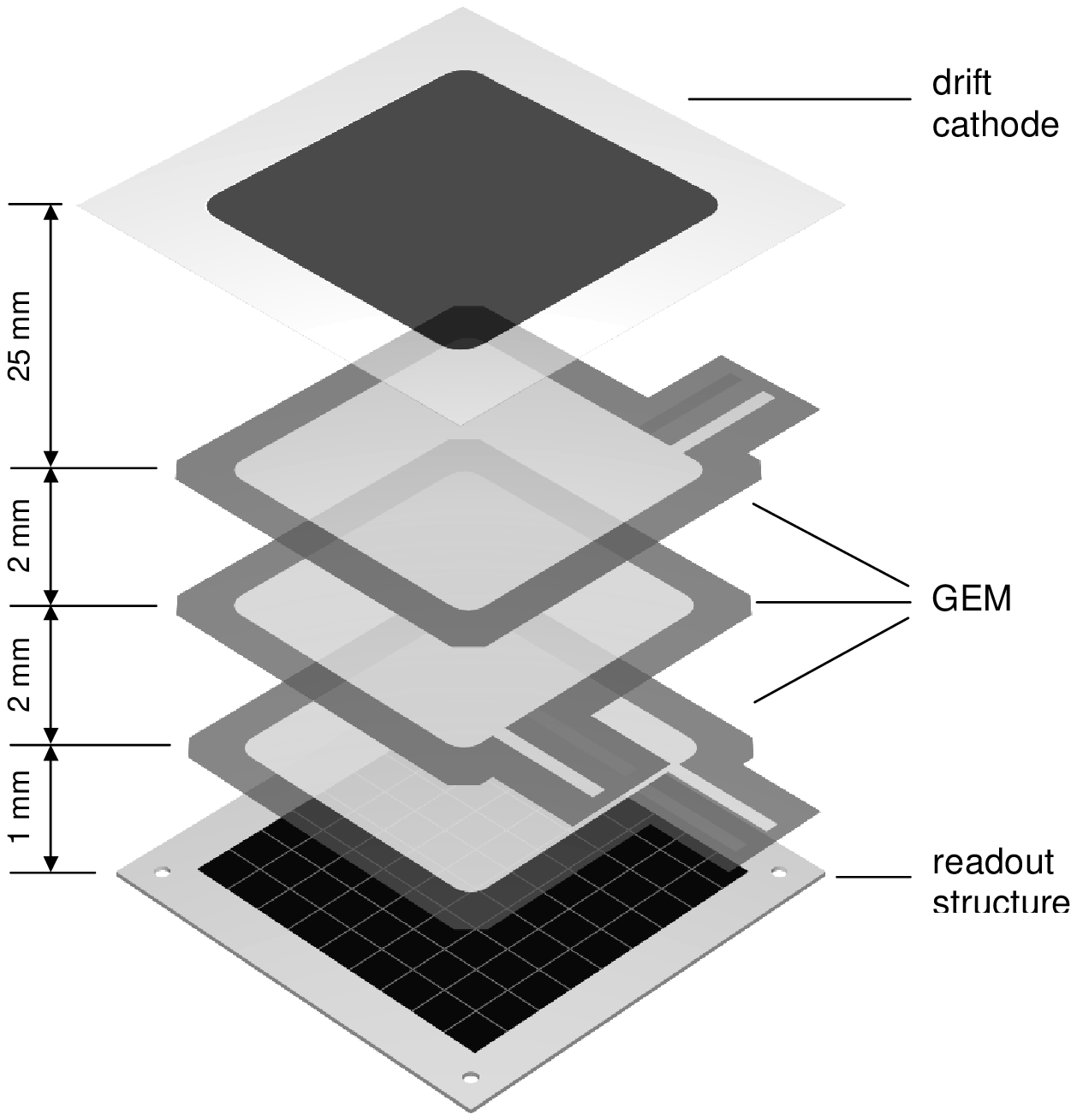}
  \caption{Schematic cross-section of the triple-GEM detector setup.
  The sensitive
  area has a size of $56\times56\,\mathrm{mm}^2$, subdivided into
  $7\times7$ cells (actually, the readout structure contains
  $9\times9$ cells, whereby the outermost cells are dummy cells).
  The drift cathode is typically supplied with a voltage of
  $-4000\,\mathrm{V}$,
  in combination with the (not shown) drift cage providing a
  homogeneous drift field of $\approx1\,\mathrm{kV/cm}$.
  The readout structure is at ground
  potential. The electric field between the individual GEMs is set
  to $2.5\,\mathrm{kV/cm}$ and between the undermost GEM and the
  readout structure to $3\,\mathrm{kV/cm}$.}
  \label{fig_schematic}
\end{figure}

Fig. \ref{fig_schematic} shows a schematic cross-section of the
detector. The X-ray photons enter the detector through the carbon
fibre entrance window and the subjacent $50\,\mathrm{\mu m}$ thick
mylar foil clad with a few micrometer thick carbon layer at the
bottom side serving as drift cathode. Inside the conversion region
between the drift cathode and the (first) gas gain structure (gap
of about $25\,\mathrm{mm}$, which is a compromise between a good
photo effect efficiency and moderate parallax) one single photon
produces a few hundred primary electrons in the gas volume, which
is typically filled with $\text{Xe/CO}_2$ (90/10).
Since this charge amount is too small to be detected additional
amplification is necessary. For that purpose micro pattern devices
like MicroCAT (micro compteur {\`a} trous) \cite{Sarvestani1998a}
or GEM \cite{Sauli} have been intensively tested
\cite{Sarvestani1999a,Orthen2002a,Orthen2003a,Orthen2003b}. The
MicroCAT has many advantages like stable gas gaining at high
rates, time stability or an enormous robustness. Nevertheless, we
were not successful in finding a suitable spacer concept which
should guarantee a constant distance to the subjacent readout
anode. Therefore, we decided not to use this device. For the GEM
an external spacer concept is not necessary and, despite some
disadvantages like rate- and time-dependent gas gaining which are
discussed in Sec. \ref{sec_performance}, we have decided to use a
constellation of three consecutive GEMs (triple-GEM).

\begin{figure}
  \includegraphics[clip,width=7cm]{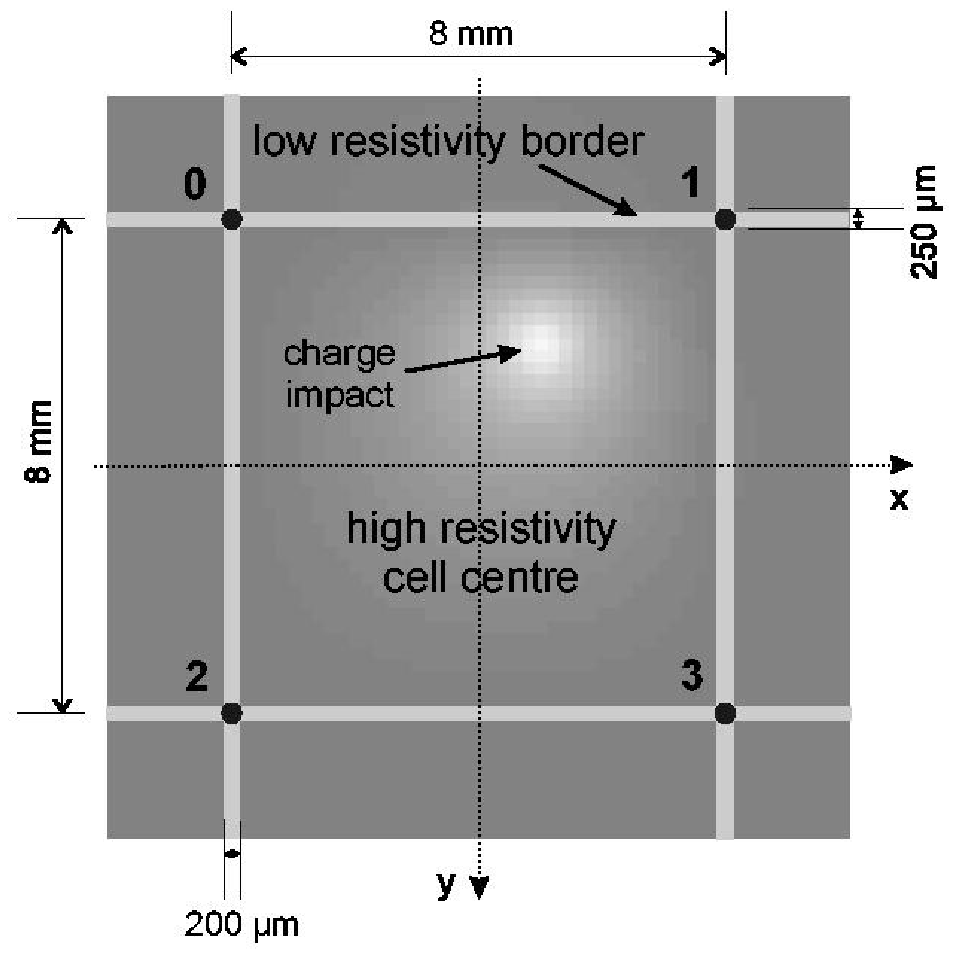}
  \caption{Schematic of one cell of the (PCB) readout structure.
  Around the charge impact point the electric potential increases,
  displayed by the white colour, and the thereby created
  electric field forces the
  charge to move towards the readout nodes at the cell corners.}
  \label{fig_rs}
\end{figure}

After multiplication the charge cluster hits the readout anode,
which is realised by an interpolating resistive structure, where
the spatial information of the event is determined in two
dimensions \cite{Besch}. The sensitive area of the readout
structure is divided into $7\times 7$ square cells. Each cell
(Fig. \ref{fig_rs}), which has an edge length of
$g=8\,\mathrm{mm}$ and a surface resistance of
$100\,\mathrm{k\Omega/ \Box}$, is surrounded by better conducting
borders with a width of about $200\,\mathrm{\mu m}$ and a surface
resistance of $1\,\mathrm{k\Omega/ \Box}$. The resistive material
is printed onto a ceramics or a PCB medium, based on FR4. The
charge flows to the readout nodes at the corners of the cells,
which are connected by small through connections to the backside
of the readout structure. By means of linear reconstruction
algorithms, e.g. using only the collected charges $Q_i$ at the
four readout nodes of one cell (4-node algorithm), the event
position can be calculated within the cell. For a proper
reconstruction using combinations of several linear algorithms
(463-node algorithm) always 9 nodes should be read out
\cite{Wagner2003}.

\begin{figure}
  \includegraphics[clip,width=8.7cm]{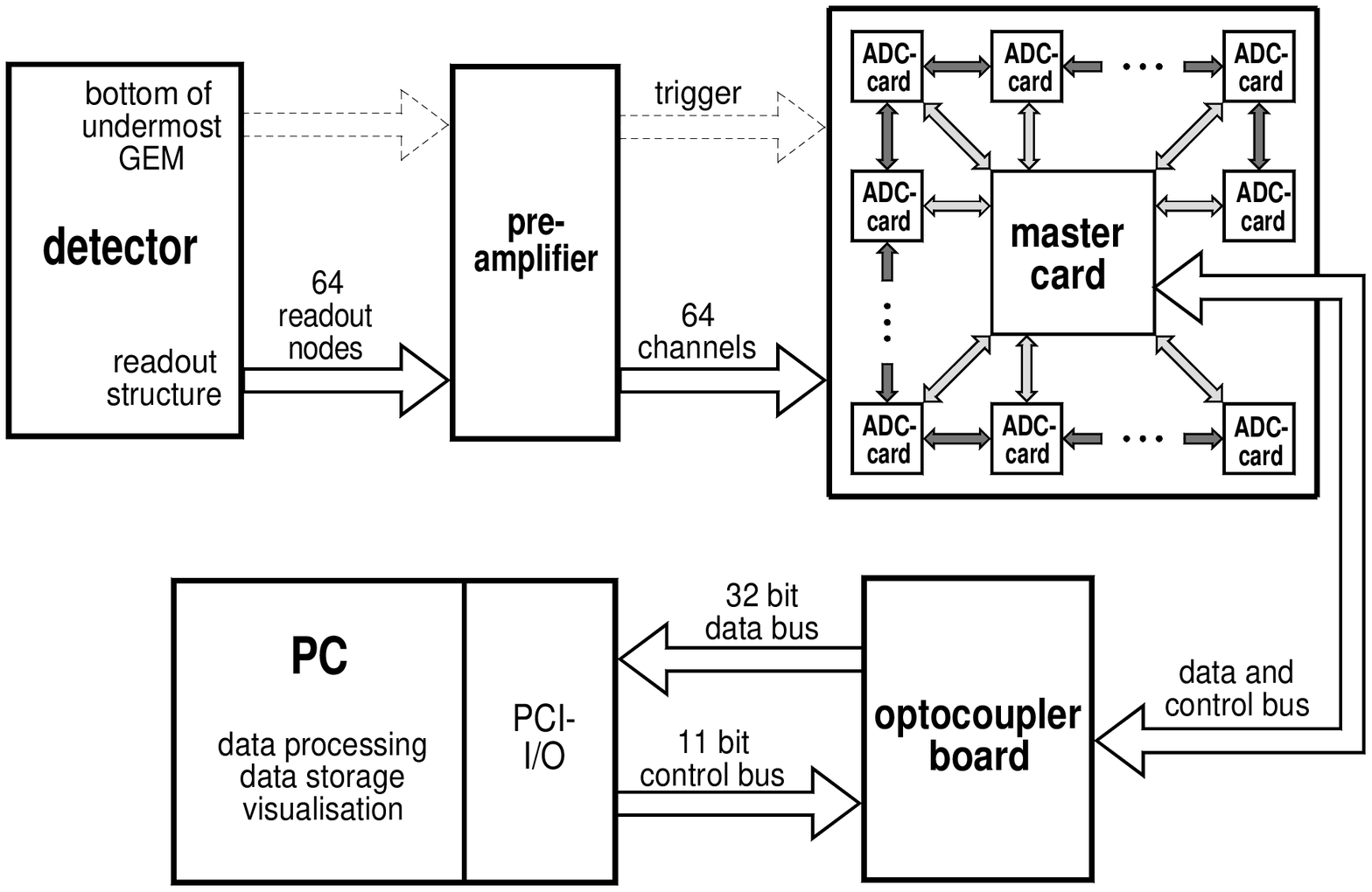}
  \caption{Block schematic of the readout.}
  \label{fig_schemel}
\end{figure}

Due to this interpolating concept the cells can be subdivided into
\emph{Virtual Pixels} (ViP) and a large sensitive area can be
covered, simultaneously reaching a spatial resolution of the order
of $100\,\mathrm{\mu m}$ (fwhm) which is nearly two orders of
magnitude better than the size of a cell (this spatial resolution
can only be obtained in combination with adequate electronics and
a gas gain of about $10^4$ since the spatial resolution is
directly proportional to the reciprocal signal to noise ratio
$\propto \text{SNR}^{-1}$).
Each read out charge is amplified by a charge sensitive amplifier
and digitized by a $12\,\mathrm{bit}$ ADC (analog to digital
converter) sampling with a frequency of $66\,\mathrm{MHz}$. Four
ADC-channels corresponding to one detector cell are placed
together at one ADC-card, where they are connected to one complex
Xilinx FPGA (field programmable gate array) logic device
\cite{Nurdan2003}. Via point to point links the sixteen ADC-cards,
containing all 64 channels, can communicate with their direct
neighbours, while the data/control bus supports the communication
with a master card, controlling the data latch to the PC (Fig.
\ref{fig_schemel}).

Together with the signal information a time stamp in terms of a
$\mathrm{\mu s}$- or $\mathrm{ns}$-counter is digitally recorded.
The counter can externally be reset. Additionally, an external
signal (e.g. a linear ramp) can be digitised by the ADC at the
master card. The time information can be used to slice the data
after the measurement has been finished.

The digitised data from the ADC-cards are stored into a FIFO
inside the Xilinx on the master card, which is connected via
optocouplers to a PC using a standard $32\,\mathrm{bit}$
PCI-interface. Currently, the transfer in DMA-mode can latch data
with a speed of $12\,\mathrm{MB/s}$ into the memory of the PC. By
using an alternative PCI-card the transfer speed can even be
increased to $80\,\mathrm{MB/s}$. A dedicated software package,
written in Visual C++, processes data online, so that the user is
directly able to evaluate the running measurements. The raw data
as well as the already processed image histogram can be stored to
hard disk. Each event occupies $25\,\mathrm{bytes}$ in the raw
data file.

\section{Detector performance \label{sec_performance}}

\subsection{Gas gain}

Each individual GEM foil produces a gas gain of about $10$--$100$,
if the GEM operation voltage amounts to about
$300$--$600\,\mathrm{V}$ depending on gas mixture and pressure.
The total gas gain of a triple-GEM constellation can easily exceed
$10^4$ and stable operation in xenon gas mixtures up to
$2\,\mathrm{bar}$ at this gain is possible \cite{Orthen2003b}.
Operation with pressure $>3\,\mathrm{bar}$ with high-$Z$ gases
like xenon is not advisable due to a high discharge probability at
gains $>5\cdot10^3$. The use of the triple-GEM for hard X-ray
detection ($E_\gamma\gtrsim25\,\mathrm{keV}$) is therefore rather
limited.

Since the GEM technology is well sophisticated and intensively
tested also large area GEMs can be built \cite{Bachmann2000} and
thus large sensitive areas are feasible.

The gain homogeneity over the total area of the triple-GEM
constellation at uniform illumination is much superior to MicroCAT
detectors because no external spacer concept is needed due to the
intrinsic constant thickness of the Kapton foil and the width of
the multiplication gap, respectively. Disadvantageous is the fact
that the GEM can easily be destroyed by too heavy sparks. To avoid
the destruction, large area GEMs should be subdivided to reduce
the capacitance of the individual sectors
\cite{Bachmann2000,Bachmann2002b}. The detector should always be
operated in safe voltage and gas gain ranges, if possible.

\begin{figure}
  \includegraphics[clip,width=7cm]{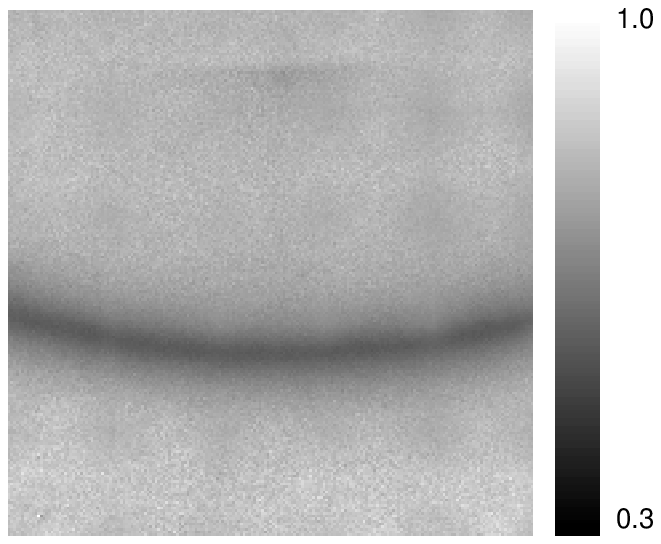}
  \caption{Relative gain in the region of the inner $5\times5$
  cells of the sensitive area during the
  measurement of a silver behenate diffraction target. In the high
  rate diffraction ring area (within the white dotted lines)
  the photon flux is about $50$ times
  higher than in the background; the gain drops roughly by
  a factor of $2$ (dark colours).}
  \label{fig_rategain}
\end{figure}

Due to charging up of the Kapton, the gas gain drops as a function
of illumination time and rate. As an example, Fig.
\ref{fig_rategain} shows the measured relative gain distribution
as a function of the position in the detector during the
measurement of a silver behenate diffraction target with high
local rates (cf. Sec. \ref{sec_agbehenate}). In counting
detectors, like the detector presented here, a rate- and
time-dependence of the gas gain is not relevant as long as the
signals do not drop under the trigger threshold. However, an
energy selectivity can not be carried out for inhomogeneous images
and due to the decreasing signal-to-noise ratio also the spatial
resolution drops in highly illuminated areas. This has to be
compensated by a very high overall gas gain (which is easily
possible with the triple-GEM) but which requires a high dynamic
range of analogue and digital electronics. Due to the time- and
rate-dependent behaviour the GEM can not be used as a
preamplification stage in integrating systems, as planned for the
MicroCAT device \cite{Menk1999}. To stop the charging of the
Kapton and thus to keep the gas gain rate- and time-independent a
coating of the GEM with amorphous carbon is proposed
\cite{Beirle}. The energy resolution of the triple-GEM then is
expected to be in the order of $20\,\%$ at $8\,\mathrm{keV}$,
which can also be obtained with standard (uncoated) GEMs in
homogeneous illumination or single photon spots with constant
X-ray flux after an equilibrium state has been reached.

\subsection{High rate capability \label{sec_hrc}}

In counting detectors one has to make a compromise between high
rate capability and spatial resolution. In general a detector can
be optimised to deal with high rates by a very fast shaping of the
analogue signals, at the same time, however, losing spatial
resolution due to too little collected charge and thus a too small
SNR. The most elegant way to avoid these problems is to produce
very short intrinsic detector signals, which is one big advantage
of micro pattern devices in contrast to MWPCs. The GEM features a
very fast signal shape at the anode, which is induced only by
electrons. The about $10^3$-times slower ions do not make any
contribution to the signal, which is however the case for other
micro pattern devices like MicroCAT or micromegas (micro mesh
gaseous structure) \cite{Giomataris1996}. In our detector setup
the raw anode signals have approximately a gaussian shape with a
mean length of $20$--$100\,\mathrm{ns}$ (fwhm) for
$\text{Ar/CO}_2$ (90/10) at standard pressure and $\text{Xe/CO}_2$
(90/10) at a pressure of $3\,\mathrm{bar}$, respectively
\cite{Orthen2003b}.

\begin{figure}
  \includegraphics[clip,width=8cm]{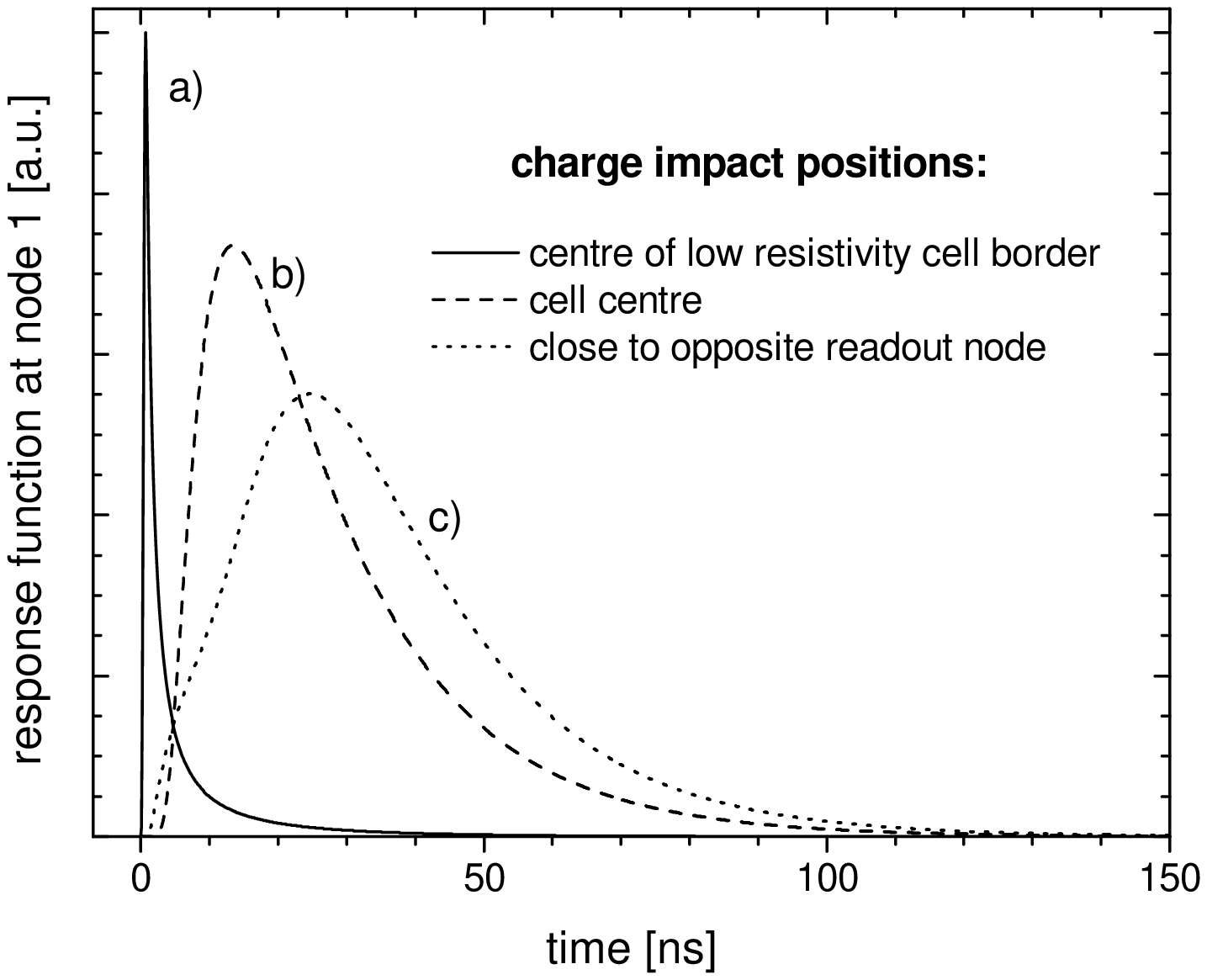}
  \caption{Simulation of the signal diffusion behaviour of the
  PCB-readout structure \cite{Wagner2002a} for a $\delta$-like input
  signal at three different charge impact positions
  (cf. Fig. \ref{fig_rs} for node denotation):
  a) symmetrically between node 0 and 1 on the low resistivity
  cell border; b) cell centre; c) close to node 2.
  The displayed currents are read out at node 1.
  Since the absolute current values
  obtained at impact positions b) and c) are
  very small compared to position a) we scaled the plotted functions
  arbitrarily to increase the figure's clearness.}
  \label{fig_ds}
\end{figure}

The relatively high resistance and capacitance of the sensitive
area leads to an integrating $RC$-element-like behaviour of the
readout structure, resulting in a temporal broadening of the input
signal (Fig. \ref{fig_ds}). After the signal information has
crossed the cell from the point of impact towards the readout
nodes, the total signal width has increased.

\begin{figure}
  \includegraphics[clip,width=18cm]{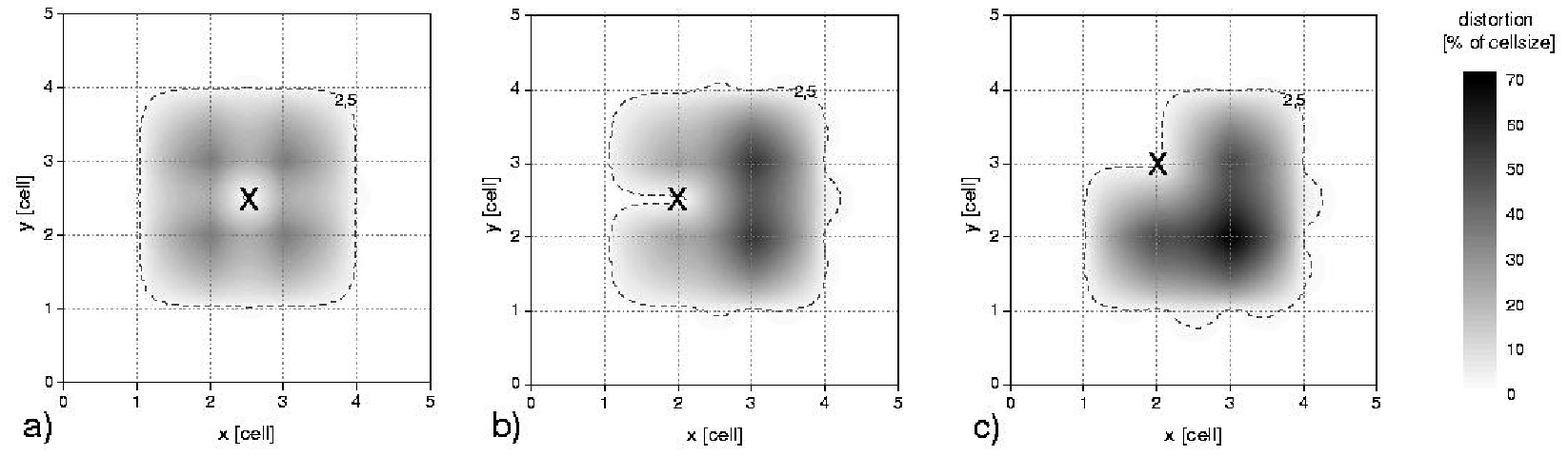}
  \caption{Worst case estimation of the reconstruction distortions
  (in percent of the cellsize), appearing
  at two contemporaneous events. The ``X" marks the impact position of
  the first event for which the distortion is determined: a) in
  the cell centre; b) at the centre of a low resistivity cell
  border; c) at a node. The absolute value of the
  expected distortions on the position of the first event ``X"
  are plotted as a function of the impact position of the second
  event. The dashed contour lines correspond to a
  distortion of $2.5\,\%$ of the cell size, i.e.
  $200\,\mathrm{\mu m}$ for a cellsize of $g=8\,\mathrm{mm}$. The ratio
  between high resistivity cell centre and low resistivity cell border
  amounts to 100.}
  \label{fig_deadarea}
\end{figure}

Due to the interpolating character of the image reconstruction we
have to make sure, that the signals do not pile up in a certain
area. Affected by this constraint is at maximum an area of
$3\times3$ cells containing the cell with the charge impact as the
central area (Fig. \ref{fig_deadarea}). In this simulation the
reconstruction is carried out with a simple 4-node algorithm,
using only the charges collected at the four central nodes. This
is a worst case estimation for two contemporaneous events. The
later the second event appears within the dead time window of the
first event the smaller is the distortion. With a typical
amplified signal having a length $<100\,\mathrm{ns}$ the average
detected flux is limited to
$10^6\,\mathrm{photons}/(3\times3\,\mathrm{cells})$ with less than
$10\,\%$ dead time loss. This value corresponds to an average
detected flux of about $2\cdot10^5\,\mathrm{photons\,cm^{-2}
\,s^{-1}}$ for a detector with a cellsize of
$8\times8\,\mathrm{mm^2}$, which is a true value. This means, that
the global count rate capability is the product of the average
detected flux and the total size of the detector. Note that apart
from pure pixel devices, a similar principle limitation in local
rate capability is found for all interpolating devices, although
it is often not explicitly mentioned. Often one finds a mismatch
between the global rate capability and the number given for the
average detected flux multiplied by the active area of the
detector. This is not the case for the numbers given in this work.

Since the arising dead area is always measured in cell units a
decrease in cell size leads to a higher average detected photon
rate per unit area. By decreasing the cell size for instance to
$2\times2\,\mathrm{mm^2}$, the average detected flux increases by
a factor of 16 to $>3\cdot10^6\,\mathrm{photons\,cm^{-2}\,
s^{-1}}$. In principle, one is free to adapt the cell size to be
able to deal with the desired photon rates. However, the cell size
is limited upwards by the deteriorating spatial resolution and the
signal diffusion, which increases nearly proportional to the
square of the cell edge length $g^2$ (this result was obtained by
simulations \cite{Wagner2004}). Decreasing the cell size is
limited by the readout speed of the electronics which can not be
exceeded. Moreover, if the cell size is chosen too small
($<1\times1\,\mathrm{mm^2}$), the advantage of the interpolating
readout system vanishes; in this case, a pure pixel device is much
more advantageous and easier to handle. By means of a complex
signal recognition double events may potentially be processed
properly in future, which would improve the local rate capability
enormously. Yet, this feature has not been investigated.

Another rather critical point in gaseous avalanche detectors is
the influence of space charge produced by the multiplication
process -- mainly of the slowly drifting ions --, causing spatial
reconstruction distortions, because the electrons, drifting from
the conversion gap towards the gain regions, are attracted by the
ion space charge. A big advantage of micro pattern devices
compared to e.g. MWPCs is the fact, that the amount of ions,
drifting back from the multiplication regions to the drift
cathode, is relatively small. In case of the triple-GEM we have
measured an ion feedback $<5\,\%$ \cite{Orthen2004}. With special
care in choosing the parameters of the GEM it should be possible
to suppress the ion feedback to less than $3\,\%$ at a gain of
$10^4$ \cite{Breskin,Bondar2003}, but this has still to be proven
experimentally. For the MicroCAT in combination with the resistive
readout structure we have measured no serious degradation in
operation due to space charge at photon peak fluxes
$>10^7\,\mathrm{photons\,mm^{-2} \,s^{-1}}$ with a beam
collimation of $\lesssim0.1\,\mathrm{mm^2}$. Also the GEM can work
with high local fluxes \cite{Ostling}. \onecolumn
\twocolumn

\subsection{Noise level and intensity precision}

The noise level of the triple-GEM detector is around a few counts
per seconds over all the sensitive area, as typical for counting
detectors. For a $280\times280$ virtual pixel interpolation,
corresponding to a pixel size of $200\times200\,\mathrm{\mu m}^2$,
the noise level is far below $10^{-4}\,\mathrm{cts\, s^{-1} \,
pixel^{-1}}$. Intensity precision is thus, as expected, only
Poisson-limited and can be better than
$1\,\text{\textperthousand}$.

\subsection{Analogue Electronics}

The charge integrating preamplifiers with subsequent shaping,
currently in use, have a shaping time of $300\,\mathrm{ns}$. They
have not been optimised for this application, but were rather
built for a MWPC which produces signals with long ion tails. In
the future we plan to use a linear transimpedance preamplifier.
The charge collection by current integration can be carried out
digitally afterwards in the FPGA. One prototype channel which has
already been tested responds to a $\delta$-like input with a
gaussian signal shape with a width of about $15\,\mathrm{ns}$
(fwhm) due to bandwidth limitations.

\subsection{Preprocessing and data readout}

The data readout crucially influences the dead time behaviour of
the overall system. In a previous version of the electronics
\cite{Stiehler1998} the data transfer used to be the bottleneck of
the readout, which limited the readout speed dramatically and
implied additionally an electronic dead time of about
$10\,\mathrm{\mu s}$ per event. The synchronous readout was
triggered by an externally generated signal (``global trigger")
which was created by the discriminated signal from the bottom side
of the lowermost GEM (or from the MicroCAT, respectively).

The newly developed very modular ADC system enables a parallel and
asynchronous readout of the nine nodes of interest, carrying the
most important information of one event. Furthermore, it offers
the capability of preprocessing before data are latched to the PC.
The processing is performed by Xilinx FPGA devices which are
freely programmable using a hardware description language like
VHDL. The new system is, for example, able to integrate the
current signal, to find the signal peak and to locate the node
with the maximum signal and the surrounding 8 nodes autonomously.
Furthermore, it can work in a self-triggering mode, i.e. a trigger
is generated by a signal recognition unit. This trigger signal
needs not serve as a global but rather as a ``local" trigger which
means that only the region of interest is triggered for readout
while the neighbouring channels (at least two cells away) are not
read out and thus work in a normal fashion. During the readout of
the 9 nodes of interest the participating ADC channels are
continuously sampling new data, so that no additional dead time
occurs. For the final version of the local trigger programming we
expect that the dead time contribution due to the readout is small
compared to the signal lengths \cite{Martoiu}.

Due to the modularity, an upgrade of to a larger system can be
realised by a replication of the system, described above. The
master cards of these sub-systems have to be connected to an
overall-master. Thereby, it should be possible to extend the
detector by scaling the sensitive area and the global rate
capability at the same time without loss of spatial resolution and
local rate capability. When the number of electronic channels
exceeds a certain limit, the whole reconstruction work should be
implemented in the FPGA logic devices on the system master-card(s)
by making use of the newest generation of FPGAs featuring fast
multiplier/divider units. By only transmitting the (already
time-sliced) histograms instead of the raw data the data transfer
to the PC will be compressed by several orders of magnitude.

Unfortunately, the measurements, described in this publication,
could only be carried out with a ``slim" version, which means that
the Xilinx devices were programmed with a less complex test mode
VHDL programme. It is still based on a global trigger which is
distributed to the whole system by one dedicated ADC-channel, that
originally used to digitise data at the corner of the sensitive
area (cf. black dead area in the lower right corner of Fig.
\ref{fig_rattail}). After the trigger has been received by the
other ADC-cards, the peak sampling point in a defined time window
of each channel is transmitted to the master card that decides
which 9 nodes are sent to the PC. The dead time of this procedure
amounts to about $4\,\mathrm{\mu s}$. The local trigger
programming is currently under development, and a simple prototype
version has already been tested successfully.

\subsection{Time-resolution \label{sec_tr}}

With the triple-GEM detector and the new digital (and analogue)
electronics it seems sensibly feasible to reach continuous
time-resolutions in the $100\,\mathrm{ns}$-range; thereby, the
time-resolution is limited by the width of the conversion gap
$d_\text{gap}$: Incoming photons may convert into electrons by
photo effect everywhere in this gap (actually more photons convert
directly behind the entrance window compared to positions directly
above the GEM due to the exponential absorption law). By this
physical process the time-resolution is limited by the maximum
electron drift time $\tau_\text{max}=d_\text{gap}/v_\text{e}$ in
the conversion gap. Typical values for the electron drift velocity
are $v_\text{e}=3\text{--}4 \,\mathrm{cm/\mu s}$ in xenon-based
gas mixtures at standard pressure \cite{Becker}. This means, that
the time-resolution is limited to about $800\,\mathrm{ns}$ for a
$25\,\mathrm{mm}$ conversion gap. The drift gap can be chosen
smaller, at the same time losing quantum efficiency. Increasing
gas pressure to increase efficiency will decrease drift velocity
and hence increase the drift time. Maybe the addition of
$\text{CF}_4$ to the xenon mixture can increase the electron drift
velocity \cite{Va'vra}. Simulations with the Magboltz programme
show promising results \cite{Magboltz2}.

The limitation of the time-resolution, described above, is
actually similar in all gaseous devices where a larger conversion
gap is needed to gain sufficient quantum efficiency. Incidentally,
the timing structure of the synchrotron beam caused by the
individual electron bunches in the storage ring is also lost due
to the slower electron drift in the conversion gap; for ELETTRA or
ESRF the minimum time difference between two bunches (with uniform
bunch pattern) amounts to about $2\,\mathrm{ns}$.

\section{Imaging performance \label{sec_imaging}}

The images, shown in the next subsections, have been recorded with
two different readout structures based either on a ceramics or on
a printed circuit board (PCB). Both substrates have advantages and
disadvantages:

The $\mathrm{AlO_2}$-ceramic medium is very planar and therefore
well suited for operation with a MicroCAT gas gain structure.
However, it suffers from a high dielectric constant
($\epsilon=9.9\,@\,1\,\mathrm{MHz}$) leading to a large
capacitance of the resistive area and therefore to a large signal
diffusion at the readout structure (cf. Sec. \ref{sec_hrc}).
Furthermore, the through-contacts have a relatively large diameter
of $300$--$400\,\mathrm{\mu m}$, leading to image distortions in
the surrounding area of the nodes.

The newly introduced PCB structure is not planar and hence not
suited in combination with the MicroCAT without a sophisticated
spacer concept. In combination with the triple-GEM, however, the
advantages outbalance the ceramics due to the lower dielectric
constant ($\epsilon=4.6\,@\,1\,\mathrm{MHz}$), the possibility to
use micro-vias at the readout nodes with a diameter of only
$200\,\mathrm{\mu m}$ and the very flexible multi-layer
configuration.

\subsection{Flatfield response}

\begin{figure}
  \includegraphics[clip,width=7cm]{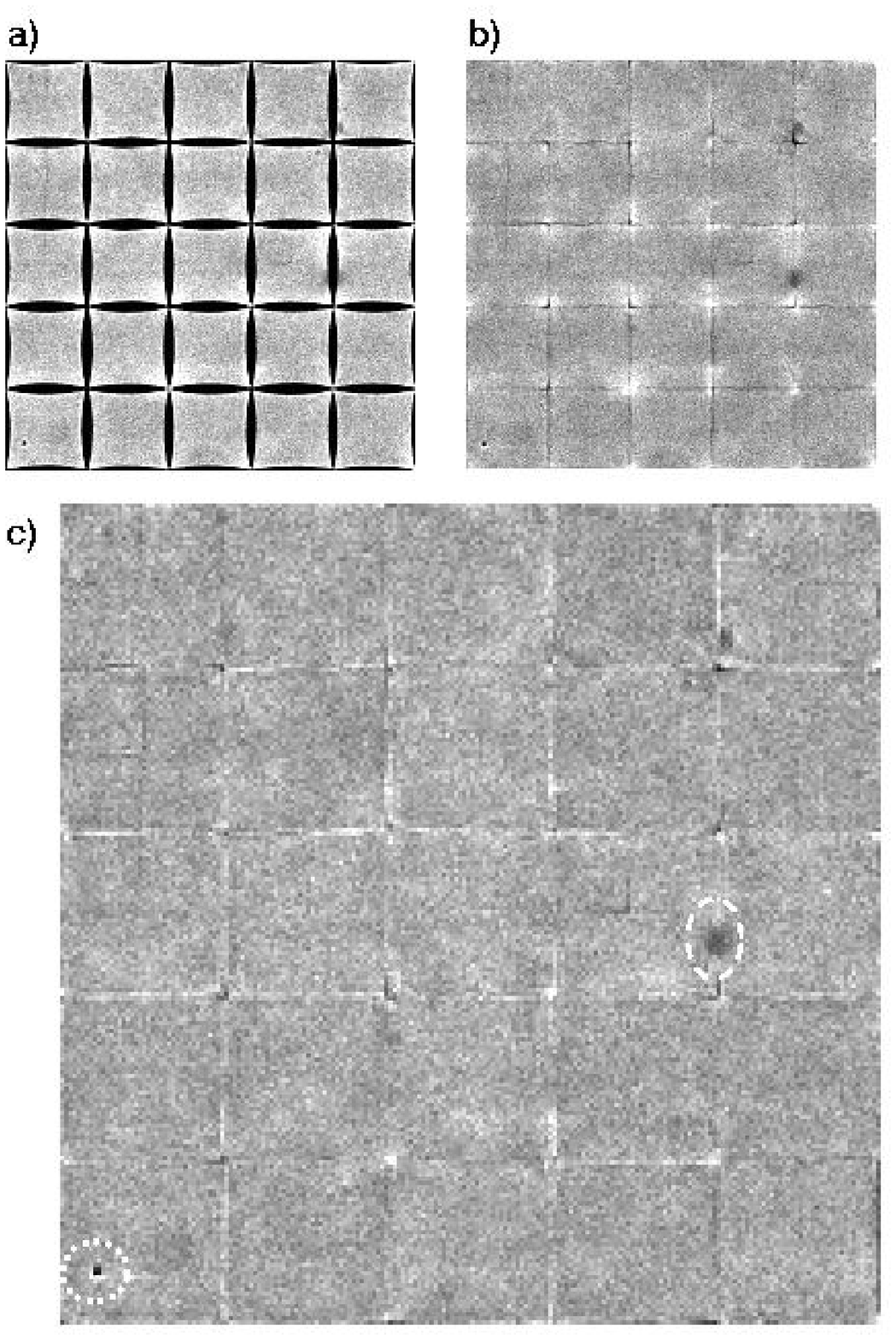}
  \caption{Reconstructed image of a flatfield
  illumination using a) the simple 4-node algorithm, b)
  the 4/6/3-node algorithm and c) the 4/6/3-node
  algorithm with subsequent non-linear correction mainly at the
  nodes and the cell borders. The flatfield shown has
  been recorded with a PCB-readout structure. Only the inner
  $5\times5$ cells corresponding to an area of
  $40\times 40\,\mathrm{mm^2}$ are depicted. A virtual
  pixel size of $200\times200\,\mathrm{\mu m}^2$ has been chosen.
  All images contain about $4.6\times10^6$ photons (mean number of
  photons per pixel $N\approx114$). The standard deviation of the
  intensity per pixel distribution amounts to $\sigma_N=54.9$ in image a),
  to $\sigma_N=16.2$ in image b) and to $\sigma_N=13.6$ in image c).
  The Poisson-limit amounts to
  $\sigma_{N-\text{Poisson}}=\sqrt{114}\approx10.7$.
  The black spot in the lower left area (dotted circle)
  is caused by a defect region in one of the three GEM structures.
  The larger dark grey spot in the right half of the image (dashed
  ellipse) can be attributed to an irregularity of the low
  resistivity cell border.}
  \label{fig_flatfield}
\end{figure}

Due to a slightly non-linear behaviour of the ViP readout
structure it is not possible to reconstruct distortion-free images
with only one linear algorithm. Fig. \ref{fig_flatfield} a) shows
the response of a detector with a PCB-readout structure to a
uniform illumination reconstructed with a simple 4-node algorithm.
A $^{55}\text{Fe}$-source has been used for illumination
($E_\gamma=5.9\,\mathrm{keV}$). The detector has been operated
with an $\text{Ar/CO}_2$ (70/30) gas filling at a pressure of
$1.2\,\mathrm{bar}$. The depletion at the cell border shows the
non-linear charge division behaviour. By using a more complex
linear 4/6/3-node algorithm \cite{Wagner2003} these kind of
distortions do not occur anymore (Fig. \ref{fig_flatfield} b)).
However, the population density is still not flat as it should be
(i.e. the nodes are overpopulated) due to systematic effects like
cross talk, electronic gain variations\footnote{Gas gain
variations displace the event position only far less than the
value of the spatial resolution \cite{Orthen2000}.}, variations of
the resistance of the sensitive area or the width of the low
resistivity cell borders or simply signal fluctuations due to
noise \cite{Wagner2004} . To further improve the image one can
apply a non-linear correction (two-dimensional variable
transformation), resulting in an image shown in Fig.
\ref{fig_flatfield} c).
All further images have been reconstructed with the linear
4/6/3-node algorithm with a global non-linear correction at the
nodes and at the cell borders.

\subsection{Aperture image}

\subsubsection{Hole grid collimator}
\begin{figure}
  \includegraphics[clip,width=6cm]{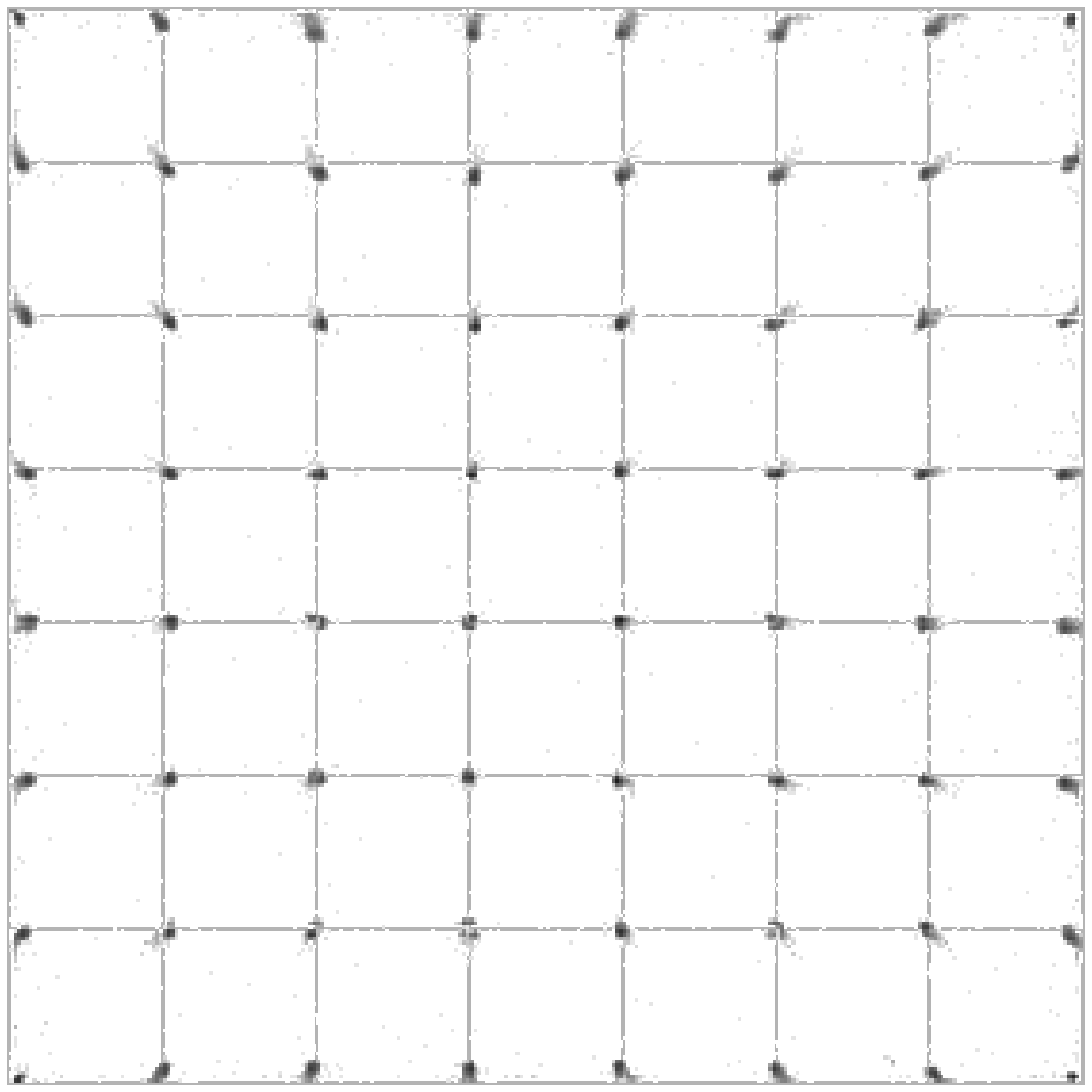}
  \caption{Image of a hole grid collimator, recorded with a
  PCB-readout structure. A
  $^{55}\text{Fe}$-source has been placed at a distance of
  $30\,\mathrm{cm}$ in front of the detector, causing
  visible parallax. The detector
  has been operated at a pressure of $1.2\,\mathrm{bar}$
  $\text{Xe/CO}_2$ (90/10) with an applied drift field of
  $1\,\mathrm{kV/cm}$.}
  \label{fig_gridcoll}
\end{figure}

The image of a hole grid collimator consisting of a stainless
steel aperture with $0.5\,\mathrm{mm}$ holes at a grid distance of
$8\,\mathrm{mm}$ is depicted in Fig. \ref{fig_gridcoll}. It
becomes obvious, that close to the border of the detector the
image is slightly distorted. These distortions are mainly caused
by a slightly inhomogeneous drift field and can be avoided by the
usage of an additional electrode frame which is fixed upon the
ceramic frame, holding the uppermost GEM. Distortions due to field
inhomogeneities can also be caused by inhomogeneous electric
fields between the individual GEMs (\emph{transfer field}) and
between undermost GEM and readout structure (\emph{induction
field}). It is very important that the distance between the
individual electrodes is as constant as possible. Especially the
transfer and induction fields are very sensitive to these
variations due to the small distance of $1$--$2\,\mathrm{mm}$ of
the particular electrodes. Transfer and induction field variations
are also a possible reason for global gas gain variations because
the effective gas gain is determined by the electron transfer
through the GEM holes which is strongly affected by the electric
fields close to the GEM structures \cite{Orthen2003a}.

\subsubsection{SAXS collimator\label{sec_saxs}}

\begin{figure}
  \includegraphics[width=14.5cm]{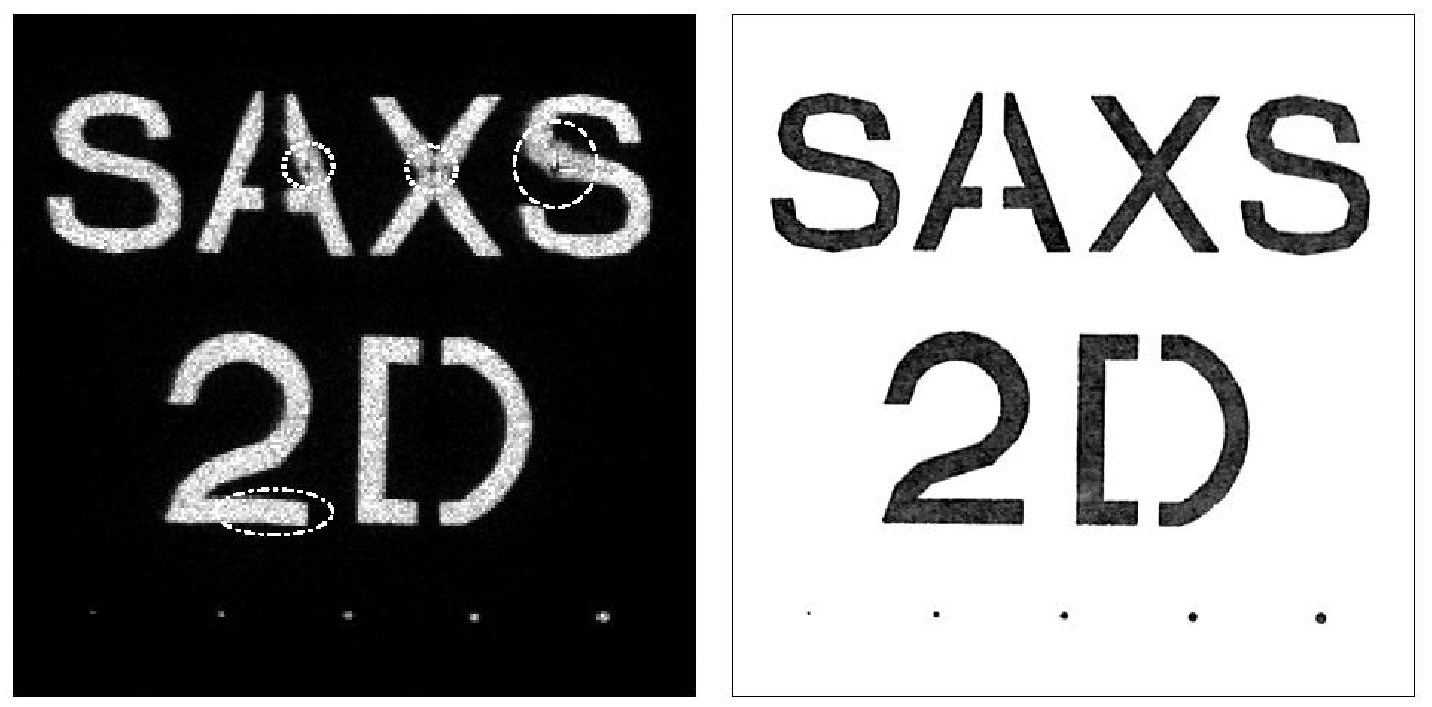}
  \caption{Image of the ``SAXS" aperture recorded with the ViP detector
  (left hand side) and photograph taken with a standard digital camera
  against the sunlight (right hand side). The detector image has
  been recorded with a beam energy of $6.4\,\mathrm{keV}$.
  The distance between detector and source amounted to about
  $2.5\,\mathrm{m}$, therefore parallax is small.
  The illuminated spots at the
  bottom correspond to holes in the aperture with diameters of
  280, 380, 480, 580 and $680\,\mathrm{\mu m}$. The aperture is
  slightly tilted by an angle of about $0.6^{\circ}$, which can
  be recognised for example by the vertical pixel jump at the
  bottom of the ``2" (dashed-dotted ellipse). The sizes of
  both depicted images amount to $4.4\times4.4\,\mathrm{cm}^2$.
  Note that the cell borders have almost disappeared.}
  \label{fig_saxs}
\end{figure}

Fig. \ref{fig_saxs} shows the image of a laser cut
$1\,\mathrm{mm}$ thick stainless steel aperture containing ``SAXS
2D" letters and five holes with increasing hole diameters. The
image has been recorded with a PCB-readout structure using photons
of an energy of $6.4\,\mathrm{keV}$ (fluorescence of a Fe-target
in a $8\,\mathrm{keV}$ synchrotron beam). The detector was filled
with a $\text{Xe/CO}_2$ (90/10) gas mixture at a pressure of
$1.3\,\mathrm{bar}$. The comparison to a photographic image of the
aperture (right hand side image in Fig. \ref{fig_saxs}) shows a
very good accordance. Only the middle part of the second ``S"
(dashed circle) looks slightly distorted; at this readout channel
the preamplifier was not working fully correctly and causes slight
distortions in all images shown. The reproduction of the small
holes shows a good agreement, too. The remaining spots (dotted
circles) are most likely due to a non-optimised flatfield
correction.\onecolumn
\twocolumn
Using the image of the smallest hole of the SAXS aperture with a
diameter of $280\,\mathrm{\mu m}$, we have determined the width of
the point spread function (psf) to be in the order of
$\sigma_\text{psf}\approx120\,\mathrm{\mu m}$ (fwhm).

\subsection{Diffraction measurement}

All diffraction patterns, presented in the following, have been
recorded at the Austrian SAXS beamline \cite{Amenitsch,Bernstorff}
at the synchrotron ELETTRA/Trieste.

\begin{figure}
  \includegraphics[clip,width=7cm]{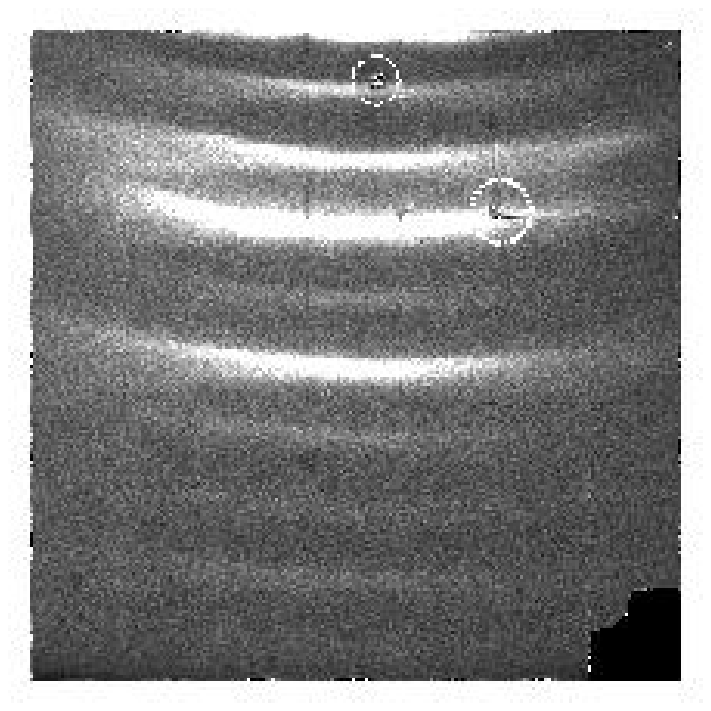}
  \caption{Diffraction image of a rat tail tendon, recorded with
  a ceramic readout structure. The image size amounts to
  $56\times56\,\mathrm{mm^2}$. The node
  at the lower right image corner was used for triggering and
  therefore caused a black dead area.
  The distorted spot in the second
  visible diffraction order (actually the $\mathrm{7^{th}}$
  diffraction order)
  is caused by a defect region in one of the
  GEM structures (dashed circle). The region of the problematic
  preamplifier channel is marked with a dotted circle.
  The image has been flatfield-corrected
  with a non optimised flatfield image.}
  \label{fig_rattail}
\end{figure}

\subsubsection{Rat tail tendon collagen}

\begin{figure}
  \includegraphics[clip,width=8cm]{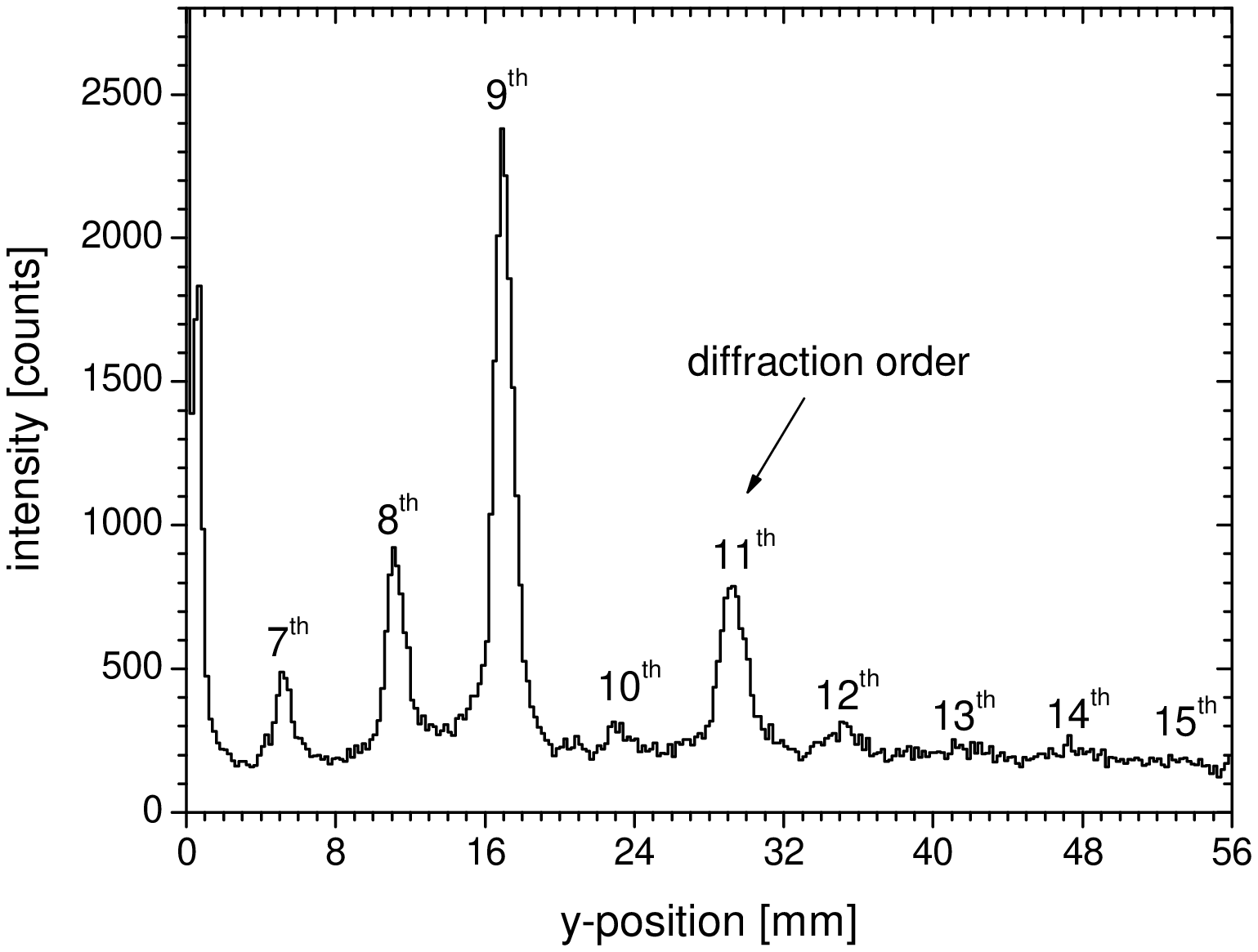}
  \caption{Intensity profile of the rat tail tendon image. 5
  pixels in $x$-direction have been summed up.}
  \label{fig_rattailprof}
\end{figure}

Fig. \ref{fig_rattail} shows the diffraction pattern of a rat tail
tendon with a $d$-spacing of about $650\,\text{{\AA}}$ which was
used for the detector calibration/alignment for the time-resolved
SAXS measurements described in Sec. \ref{sec_TSAXS} The image has
been recorded with a ceramic readout structure and a beam energy
of $8\,\mathrm{keV}$; the remaining detector parameters were the
same as described in Sec. \ref{sec_saxs} The vertical intensity
profile is shown in Fig. \ref{fig_rattailprof}. Even the
$\mathrm{14^{th}}$ and $\mathrm{15^{th}}$ diffraction order are
slightly visible.

\subsubsection{Silver behenate \label{sec_agbehenate}}

\begin{figure}
  \includegraphics[clip,width=7.5cm]{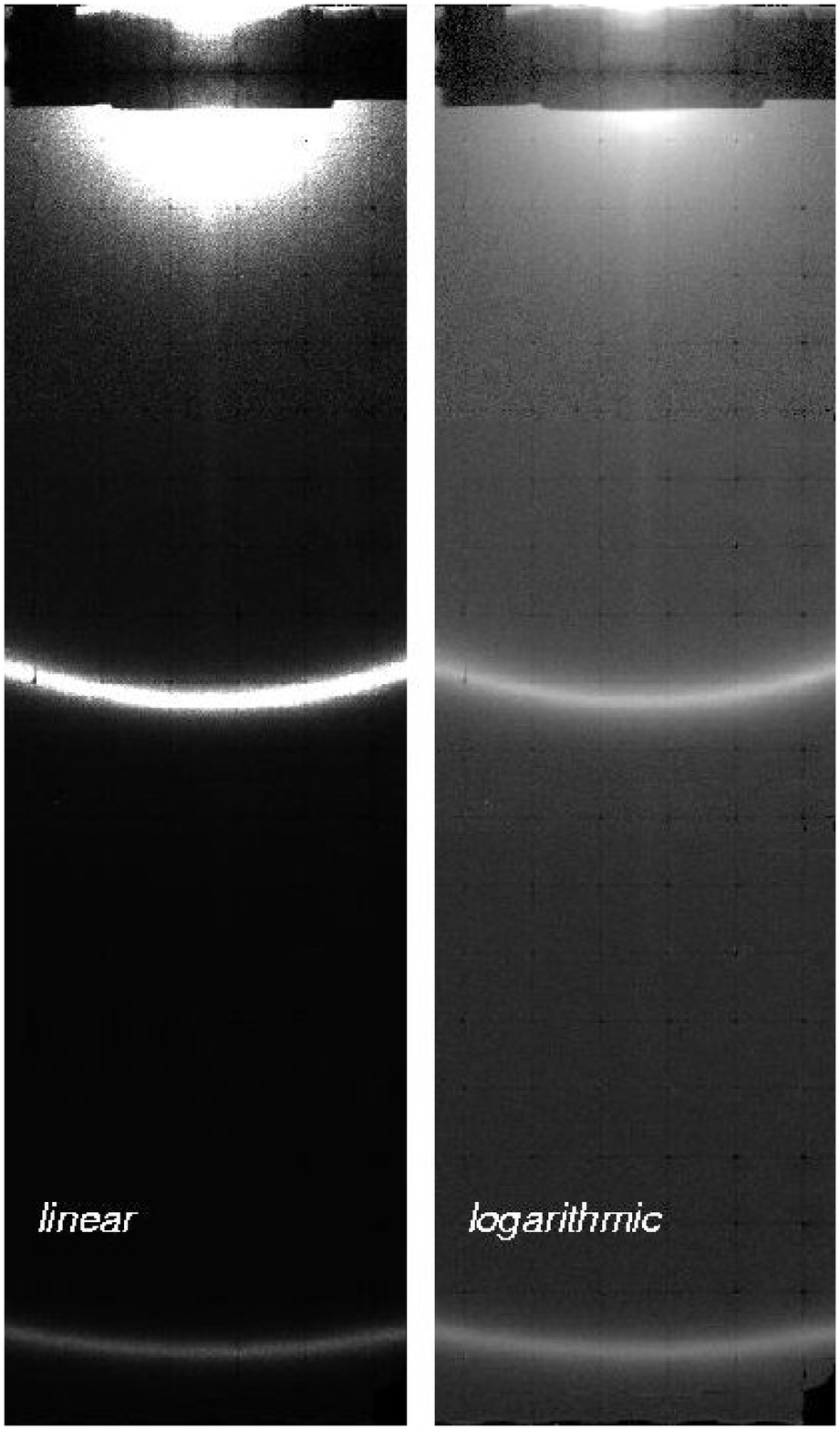}
  \caption{$1\times4$-scan of a silver behenate diffraction
  image both with linear and logarithmic scaling. One depicted image
  has a size of $48\times168\,\mathrm{mm}^2$.
  The single images, recorded with a PCB-readout structure,
  have been flatfield-corrected with a non-optimised flatfield image.}
  \label{fig_agbehenate}
\end{figure}

Another standard diffraction target in the small angle region,
silver behenate $(\text{CH}_3(\text{CH}_2)_{20}\text{--COOAg})$
with a $d$-spacing of $d_{001}=58.38\,\text{\AA}$ \cite{Blanton},
has also been recorded with the ViP detector system. The depicted
image, composed of $1\times4$ individual images, is shown in Fig.
\ref{fig_agbehenate}.

\section{Time-resolved SAXS \label{sec_TSAXS}}

Several mechanical test measurements in our laboratory have been
carried out with a X-ray tube showing the capability of the
detector system to resolve very fast repetitive processes up to
the $\mathrm{\mu s}$-regime \cite{Sarvestani2001}. Since a slower
electronic system has been used for these measurements the
processes had to be periodically repeated a few hundred or even a
few thousand times to collect sufficient statistics. A repetition
is no problem at radiation stable materials which have been used
for these measurements. However, the number of repetitions in e.g.
biological X-ray diffraction measurements is limited due to severe
radiation damage of the sample.

Here, we present a time-resolved temperature jump experiment of a
1-palmitoyl-2-oleoyl-\emph{sn}-phosphatidylethanolamine (POPE)
lipid. The sample preparation is described elsewhere
\cite{Rappolt2003}. The sample has been placed inside a
thin-walled quartz capillary with a diameter of $1\,\mathrm{mm}$.
The $T$-jump has been induced by a light flash from a solid state
laser at a wavelength of $1540\,\mathrm{nm}$ with erbium in glass
as the active laser medium \cite{Rapp1991}. We have repeated the
experiment about ten times, which was sufficient due to the high
scattering power of the sample and since we were not interested in
intensity resolution but rather in the change of the $d$-spacing
in the lipid due to the rapid temperature increase, manifested by
a spatial displacement of the diffraction ring.

\begin{figure}
 \includegraphics[clip,width=8.7cm]{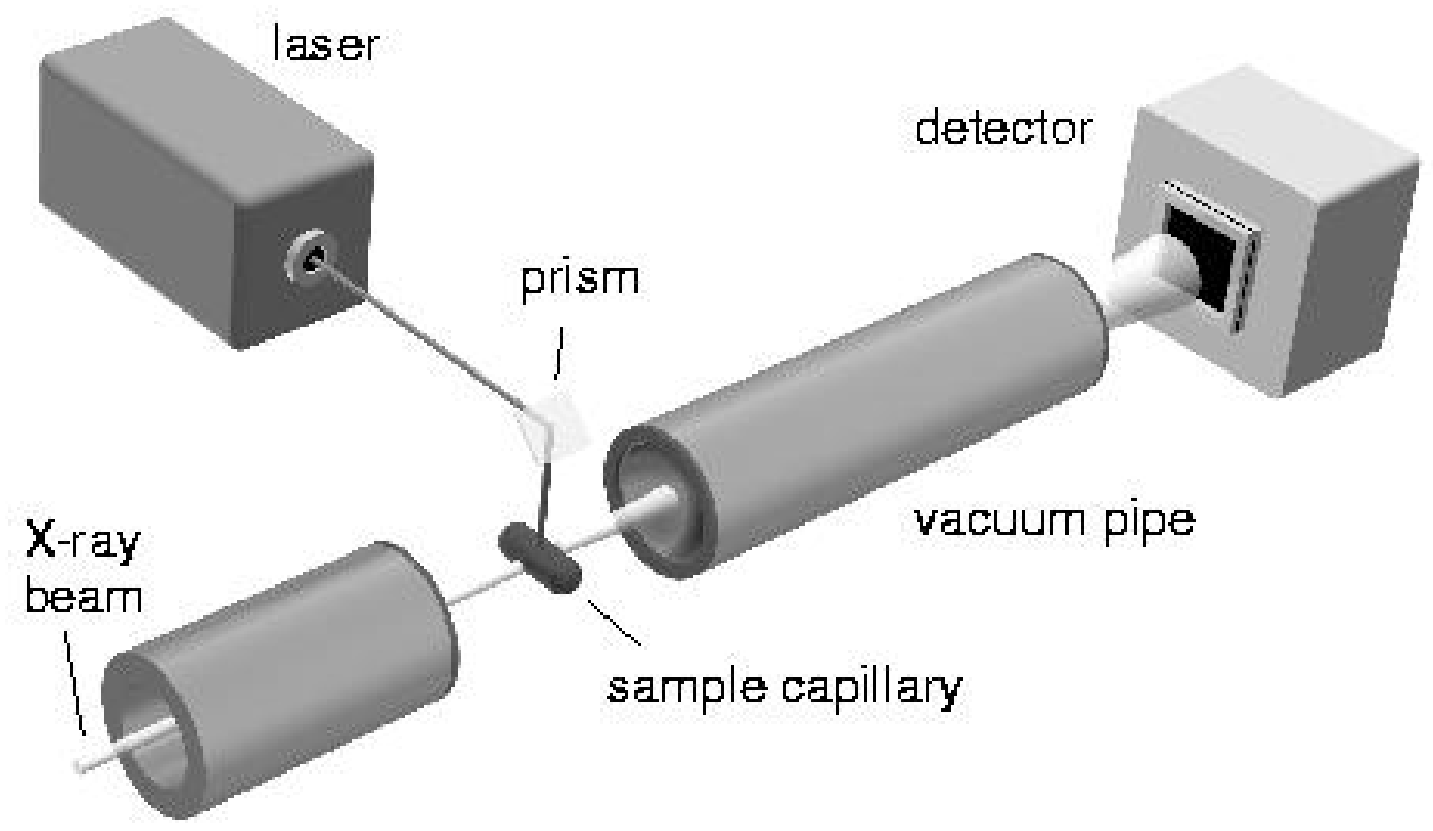}
 \caption{Schematic setup of the $T$-jump experiment at the SAXS
  beamline.}
  \label{fig_laserexp}
\end{figure}

The sketch of the experimental setup is depicted in Fig.
\ref{fig_laserexp}. The prepared lipid sample in a capillary
device has been mounted at the intersection point of the infrared
laser beam and the monochromatic $8\,\mathrm{keV}$ X-ray beam. A
water-based cooling system combined with a heater (KHR, Anton
Paar, Graz, Austria) assured a constant starting temperature of
the sample which can be chosen within a temperature range of
$0$--$150\,^\circ\mathrm{C}$. The sample temperature has been
measured with a platinum resistor PT-100. Before the $T$-jump, the
sample was equilibrated for a period of about $10\,\mathrm{min}$
at a fixed temperature. The flash-like laser pulse with a length
of $2\,\mathrm{ms}$ and a measured pulse energy of
$0.64\,\mathrm{J}$ appears about $1\,\mathrm{ms}$ after the
trigger pulse and heats the sample very rapidly by approximately
$5\,\mathrm{K}$. We have collected data for a period of up to
$30\,\mathrm{s}$ starting about $1\,\mathrm{s}$ before triggering
of the laser pulse.

\begin{figure}
 \includegraphics[clip,width=8cm]{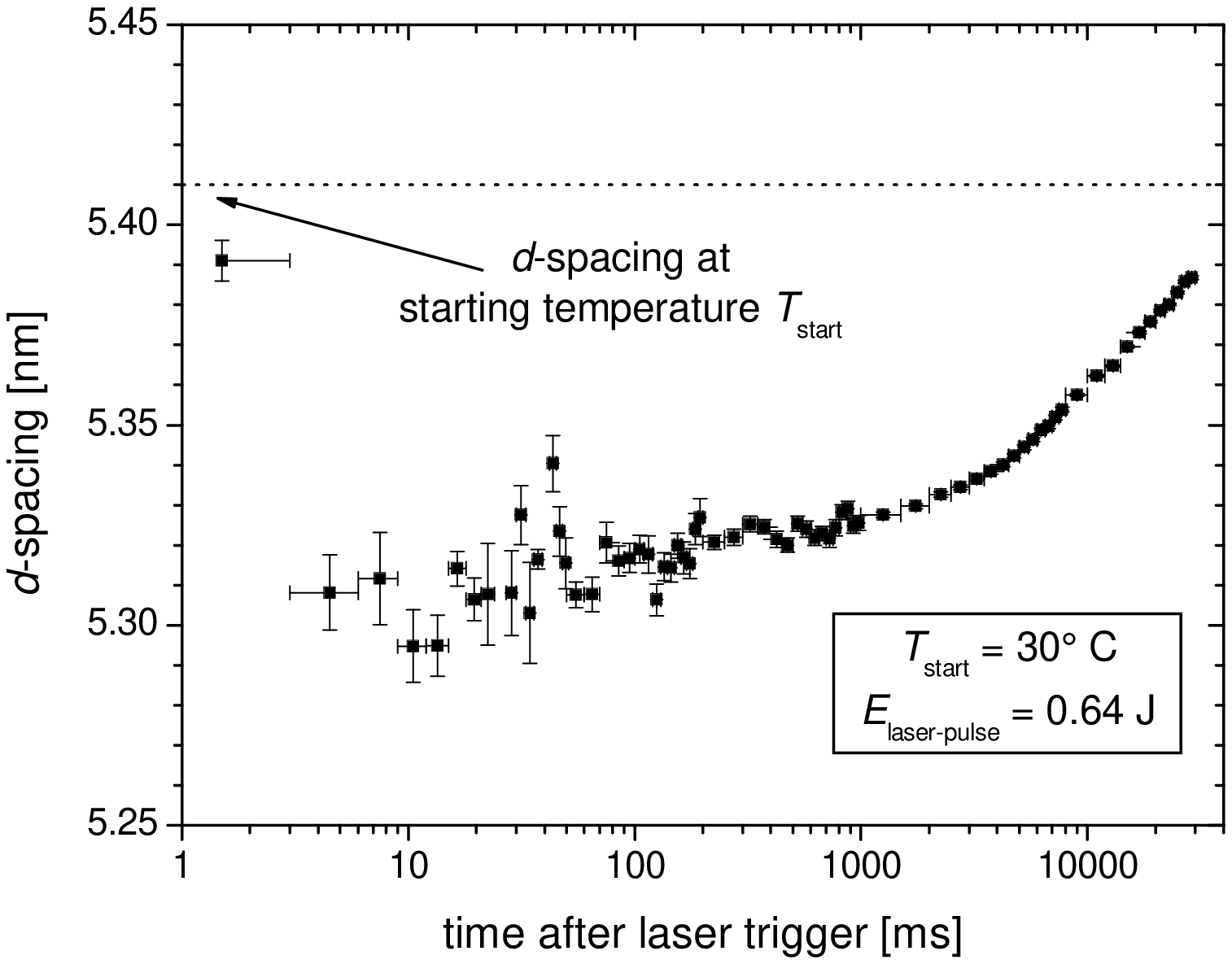}
  \caption{Measured $d$-spacing of the POPE lipid up to
  $30\,\mathrm{s}$ after the trigger of the
  laser pulse (at $t=0$). The applied
  time slices have variable widths to give an overview over 4 orders
  of magnitude in time. The bars in $t$-direction mark the
  slice-windows. Very fine $3\,\mathrm{ms}$-slices directly
  after the laser trigger show that the maximum change in $d$-spacing
  (and hence the maximum temperature change) occurs immediately
  after the end of the laser pulse (i.e. $3$--$30\,\mathrm{ms}$).}
  \label{fig_lipidtemp}
\end{figure}
After the measurement the data have been cut into fine time slices
(e.g. $3\,\mathrm{ms}$-slices for $0\le t\le60\,\mathrm{ms}$,
where $t$ denotes the time after the laser trigger). For the data
analysis we have carried out a radial integration of the
two-dimensional intensity histogram (the rat tail tendon image was
used for the determination of the beam centre at the detector) and
a subsequent lorentzian maximum-likelihood fit to the peak. The
result of the reconstructed lipid $d$-spacing as a function of
time after the laser trigger of one recorded $T$-jump is shown in
Fig. \ref{fig_lipidtemp}. It is clearly visible that the
temperature increase of the lipid features a decrease in
$d$-spacing. More detailed results concerning the $T$-jump of the
POPE lipid will be published later.

\section{Conclusion}

The GEM-based ViP photon counting prototype detector represents a
powerful tool for time-resolved X-ray imaging applications in the
$\mathrm{ms}$ or even the $\mathrm{\mu s}$-range
\cite{Sarvestani2001}. The most important parameters are
summarised in Tab. \ref{tab_comparison}, where also a possible
future version of the detector and a comparison to the RAPID
detector is listed. Large sensitive detection areas in combination
with spatial resolution in the order of $100\,\mathrm{\mu m}$ are
feasible with the virtual pixel readout concept, at the same time
saving expenses due to an enormous reduction of electronic
channels. Since the GEM produces short signals the detector can
deal with high local rates without signal pile-up; average
detected fluxes of $>2\cdot10^5\,\mathrm{photons\,cm^{-2}\,
s^{-1}}$ with less than $10\,\%$ dead time losses are easily
possible. The rejection of multiple events in an area of
approximately $3\times3$ cells around the position of the photon
impact could possibly be overcome in future by a more complex
signal processing. It is expected, that the electronic local
trigger concept -- enabling an asynchronous readout -- will be
capable of dealing with very high rates so that the global rate
capability is proportional to the number of cells and the
sensitive area, respectively; nevertheless, this has still to be
tested. The ViP detector offers a high quantum efficiency and a
relatively good parallax suppression even for higher photon
energies up to about $25\,\mathrm{keV}$ since high pressure
operation up to 2--$3\,\mathrm{bar}$ in Xenon-mixtures is
possible. The overall time resolution, which is mainly determined
by the electron drift time in the conversion gap, is at minimum in
the order of a few $100\,\mathrm{ns}$ and thus in the
sub-$\mathrm{\mu s}$-range.

\onecolumn
\begin{table}
\caption{Specification of the present and a possible future ViP
detector and comparison to the RAPID system
\cite{Lewis1997,Lewis2000}. $^{(\dagger)}$Average detected flux
multiplied by detector area (proportional by design).
$^{(\ddagger)}$Currently limited by global trigger technique.
$^{(\sharp)}$Independent of size. $^{(\Diamond)}$Limited by drift
time. $^{(\Box)}$Limited by memory organisation.
$^{(\star)}$Expected with coating of the GEMs.}
\label{tab_comparison}
\begin{tabular}{llll}      % Alignment for each cell: l=left, c=center, r=right
                        & ViP (present) & ViP (future)  & RAPID \\
\hline
 Detector type          & Triple-GEM  & Triple-GEM & Wire microgap \\
 Active area            & $5.6\times5.6\,\mathrm{cm}^2$ & $20\times20\,\mathrm{cm}^2$ & $20\times20\,\mathrm{cm}^2$ \\
 No. of cells           & $7\times7$ & $50\times50$ & - \\
 Cell size              & $8\times8\,\mathrm{mm^2}$ & $4\times4\,\mathrm{mm^2}$ & - \\
 Detected peak flux     & $>10^7\,\mathrm{photons\,mm^{-2}\,s^{-1}}$ & $>10^7\,\mathrm{photons\,mm^{-2}\,s^{-1}}$ & $>10^6\,\mathrm{photons\,mm^{-2}\,s^{-1}}$ \\
 Average detected flux  & $>2\cdot10^5\,\mathrm{photons\,cm^{-2}\,s^{-1}}$ & $>8\cdot10^5\,\mathrm{photons\,cm^{-2}\,s^{-1}}$ & $>4\cdot10^4\,\mathrm{photons\,cm^{-2}\,s^{-1}}$ \\
 Global counting rate   & $>10^6\,\mathrm{photons\,s^{-1}}^{(\dagger,\ddagger)}$ & $>3\times10^8\,\mathrm{photons\,s^{-1}}^{(\dagger)}$ & $>1.5\times10^7\,\mathrm{photons\,s^{-1}}^{(\sharp)}$ \\
 Gas filling            & $1.3\,\mathrm{bar}$ $\text{Xe/CO}_2$ & $2\,\mathrm{bar}$ $\mathrm{Xe/CO_2/CF_4}$ & $\text{Xe/Ar/CO}_2$ \\
 Conversion gap         & $25\,\mathrm{mm}$ & $8\,\mathrm{mm}$ & $15\,\mathrm{mm}$ \\
 Efficiency
 $@\,8\,\mathrm{keV}$     & $100\,\%$ & $90\,\%$ & $70\,\%$ \\
 No. of pixels          & adj., typ. $280\times280$ & adj., typ. $2000\times2000$ & adj., typ. $1024\times1024$ \\
 No. of electronic
 channels               & $8\times8$ & $50\times50$ & $128\times128$ \\
 Spatial resolution     & $<150\,\mathrm{\mu m}$ fwhm & $<100\,\mathrm{\mu m}$ fwhm & $\approx300\,\mathrm{\mu m}$ fwhm \\
 Noise level            & $\approx2.5\times10^{-4}\,\mathrm{cts\,mm^{-2}\,s^{-1}}$ & $\approx2.5\times10^{-4}\,\mathrm{cts\,mm^{-2}\,s^{-1}}$ & $\approx2.5\times10^{-4}\,\mathrm{cts\,mm^{-2}\,s^{-1}}$ \\
 Time resolution        & $<600\,\mathrm{ns}^{(\Diamond)}$ & $<250\,\mathrm{ns}^{(\Diamond)}$ & $10\,\mathrm{ms}^{(\Box)}$\\
 Spectral resolution
 $@\,8\,\mathrm{keV}$   & sufficient for triggering & $20\,\%^{(\star)}$ & $20\,\%$ \\
\end{tabular}
\end{table}
\twocolumn

\ack{Acknowledgements}

We gratefully acknowledge the help of D. Junge who assisted in
equipping the new electronics. We thank the inner tracker group of
the HERA-B collaboration for providing several GEM foils.

This work has been supported by the European Community (contract
no. ERBFMGECT980104).

\referencelist[reflist]

\end{document}